# High Efficiency, Low Cost, RF Sources for Accelerators and Colliders

R. Lawrence Ives[1], Michael Read[1], Thuc Bui[1], David Marsden[1], George Collins[1], Brian Chase[4], John Reid[4], Chris Walker[3], Jeff Conant[3], Ricky Ho[2], Jim Potter[5], Leroy Higgins[2], Aaron Jensen[6], and Henry Freund[1]

[1]Calabazas Creek Research, Inc., 690 Port Drive, San Mateo, CA 94404
[2]Communications & Power Industries, LLC, 811 Hansen Way, Palo Alto, CA 94304
[3]Communications & Power Industries, LLC, Beverley Microwave Division, Beverley, MA
[4]Fermi National Laboratory, Batavia, IL, 60510
[5]JP Accelerator Works, Inc., 2245 47th Street, Los Alamos, NM 87544
[6]Leidos, Inc., 11951 Freedom Drive, Reston, VA 20190

## Executive Summary

Numerous U.S. Department of Energy (DOE) workshops and studies describe the aspirations of the high energy physics community in addressing the technical requirements and applications for accelerators and colliders [1-4]. These range from advancing understanding of the universe to purifying water. While the energy required for these applications spans orders of magnitude, a common requirement is to reduce cost and increase efficiency of RF sources. This is most apparent for large systems requiring hundreds or thousands of sources where acquisition costs can be prohibitive, or CW systems where efficiency is of paramount importance. For many applications, size, weight, and input power can be critical. Once electrons are emitted at the input to an accelerator, it is the RF source(s) that determines the final energy at the ultimate destination.

Calabazas Creek Research, Inc. (CCR) and its collaborators are developing high efficiency, low-cost RF sources operating from a few hundred MHz to X-Band and power levels from tens to hundreds of kilowatts with the goal of providing MW-relevant sources. The efficiencies exceed 80% with projected costs as low as $0.50/Watt. Successful development of these sources will significantly alter the cost/performance landscape for RF power generation.

The RF sources under development include phase-locked magnetrons with amplitude control, multiple beam triodes, high efficiency klystrons, and multiple beam IOTs. A prototype magnetron system was built and successfully tested, with efficiency exceeding 80% under all operating conditions. The high efficiency klystron is nearing final assembly with high power tests scheduled for summer 2022. The multiple beam triode-based RF sources are also under assembly, with the first tube nearing completion. Tests are scheduled for late spring into summer 2022. The multiple beam IOT final design is nearing completion, and parts procurement is beginning. Tests will likely begin in early 2023.

These RF sources span the frequency range from 300 MHz to 1.3 GHz, and each is targeted toward new innovations that increase the efficiency and significantly reduce cost. All sources are targeted toward power levels exceeding 100 kW. The magnetron and triode programs are specifically targeted toward reduced acquisition and operating costs. The efficiency goal for each program exceeds 80% with the magnetron system acquisition cost approximately $1/Watt and the triode-based RF source at $0.50/Watt. The prototype magnetron system demonstrated both the price and efficiency goal.

The klystron is using the Core Oscillation Method (COM) to demonstrate that efficiencies exceeding 80% can be achieved in a klystron. If successful, the approach will be applicable to klystrons at different powers and frequencies. The multiple beam IOT program is investigating the impact of adding the 3rd harmonic to the drive power to increase efficiency. The program is also implementing a simpler, lower cost input drive line and moly grids to reduce cost.

These programs are exploring new approaches toward high power RF generation to increase efficiency and reduce cost. It is anticipated that completion of these programs will provide new information and innovations to further RF source technology.

# Introduction

This white paper describes four research programs at Calabazas Creek Research, Inc. focused on increasing the efficiency and reducing the acquisition cost of high-power RF sources. These programs involve collaborators from national laboratories, research companies, and RF source manufacturers. The sources are targeted toward applications requiring more than 100 kW of high duty RF power and cover the frequency band from 300 MHz to 1.3 GHz with potential applications at higher frequencies. The research is utilizing power grid tubes, magnetrons, inductive output tubes (IOTs), and klystrons; exploring innovative designs and operating configurations for each focused on efficiency and cost. It is anticipated that completion of these program will provide new information and innovations to advance RF source technology.

Each RF source investigated provides different capabilities and limitations. Power grid tubes offer the lowest cost per Watt, but are limited in gain, as are IOTs. Magnetrons are also relatively low cost but are limited in output power and lifetime. Klystrons offer increased flexibility for gain, power, bandwidth, and noise, but are relatively expensive, particularly when compared to grid tubes, magnetrons, and IOTs. Consequently, each source type is used for specific applications where it's most ideally suited. All these sources are widely used in industrial, medical, defense, and scientific applications. Consequently, completion of this research will impact a broad range of applications.

## Phase and Amplitude Controlled Magnetrons

The magnetron is a highly efficient and relatively inexpensive source of RF power. Magnetrons providing 100 kW with efficiencies exceeding 85% are available at 915 MHz and are commonly used in industrial RF heating systems. These are free-running oscillators and not suitable for systems requiring control of the amplitude and phase, including many accelerator systems. While phase locking of magnetrons is well established, amplitude control on a fast time scale is required for many accelerator applications. CCR, Fermilab and Communications & Power Industries, LLC (CPI) developed a 1.3 GHz, magnetron-based, RF system for these accelerator applications [5,6].

In this program, CPI scaled a 915 MHz magnetron to produce 100 kW at 10% duty at 1.3 GHz. Amplitude control uses a technique developed at Fermilab that employs phase modulation of the locking signal [7]. Phase modulation shifts RF power to sidebands that are rejected by high Q loads, such as superconducting accelerator cavities. The power in the sidebands represents a reduction in the power delivered to the cavity and provides amplitude control on very fast time scales.

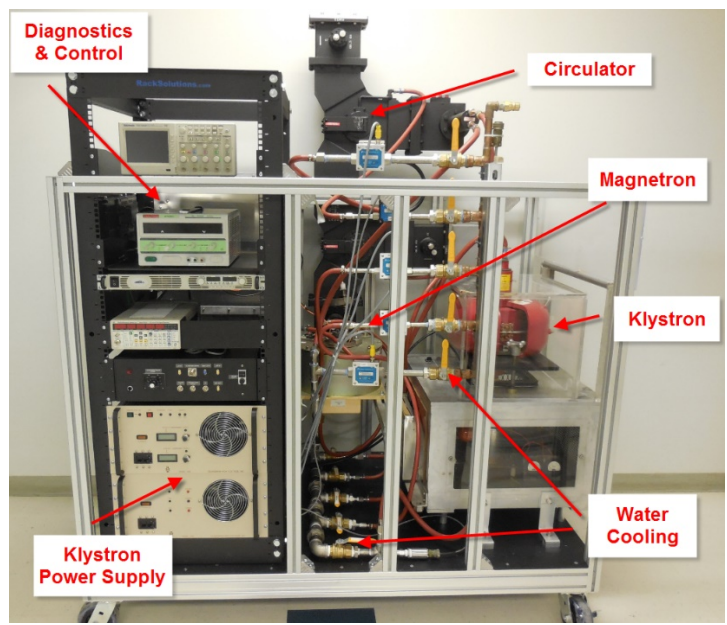

Figure 1. Magnetron system.

The magnetron was integrated into a transportable system, shown in Figure 1. It included a 5 kW klystron driver for phase locking, circulator, klystron power supply, magnetron solenoid power supply, cooling manifolds, system interlocks, and control and diagnostic electronics. Only the high voltage, pulsed power for the magnetron was external to the system. A solid-state RF source providing approximately 350W can replace the klystron for phase locking.

Figure 2 shows the performance as a free running oscillator, which is typical for magnetrons. The magnetron achieved the design output power of 100 kW. The pulse was 1.5 milliseconds at a pulse repetition rate, limited only by availability of processing time, of 2 Hz. The efficiency varied from 80% - 85% over the operating range.

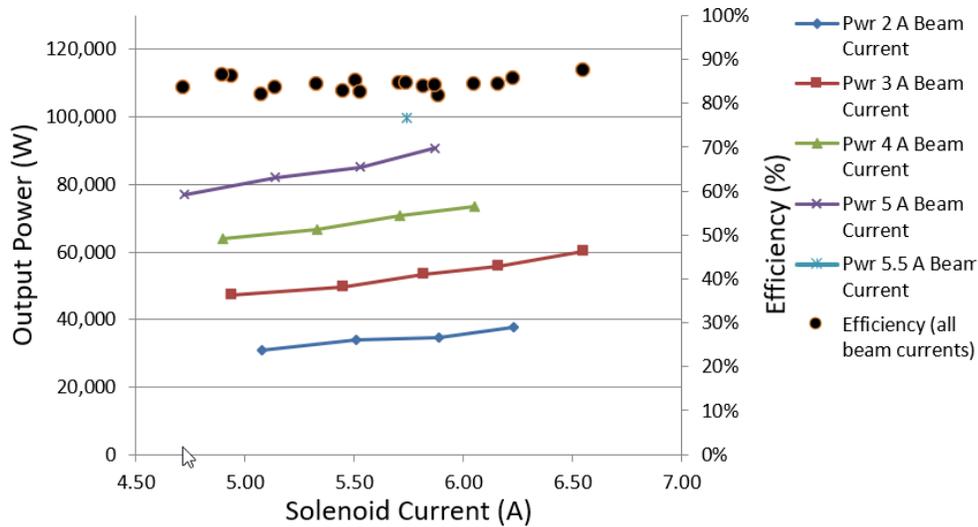

**Figure 2. Magnetron output power and efficiency as a function of solenoid current. The efficiency is almost constant at 80% - 85% throughout the operating range.**

Spectra for the magnetron in free running and phase locked modes are shown in Figure 3. Locking was confirmed by observation of the interference frequency (IF) signal from a balanced mixer. The local oscillator (LO) and RF signals were provided by the klystron driver and the sampled output from the magnetron. When the magnetron was locked, the IF signal was constant during the pulse. The first side lobes were reduced by about 5 dB in the locked mode.

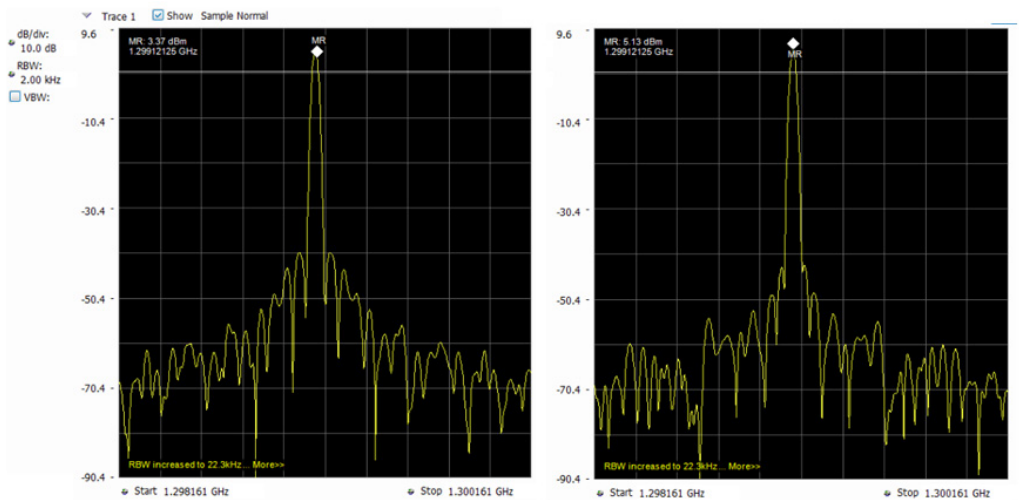

**Figure 3. Left: Free running magnetron; Right: Driven magnetron at -15 dB down at natural frequency of the magnetron.**

Figure 4 shows the effect of phase modulation of the locking signal on the magnetron output. The magnetron operated at essentially the natural (unlocked) frequency when locked with a signal 25 dB below the output power. Phase modulation was introduced via the low-level oscillator driving the klystron at 50 kHz. The figure clearly shows power diverted to the side bands as the phase modulation increases, reducing power at the center frequency. This would result in reduction of power coupled into a high Q cavity on a fast time scale.

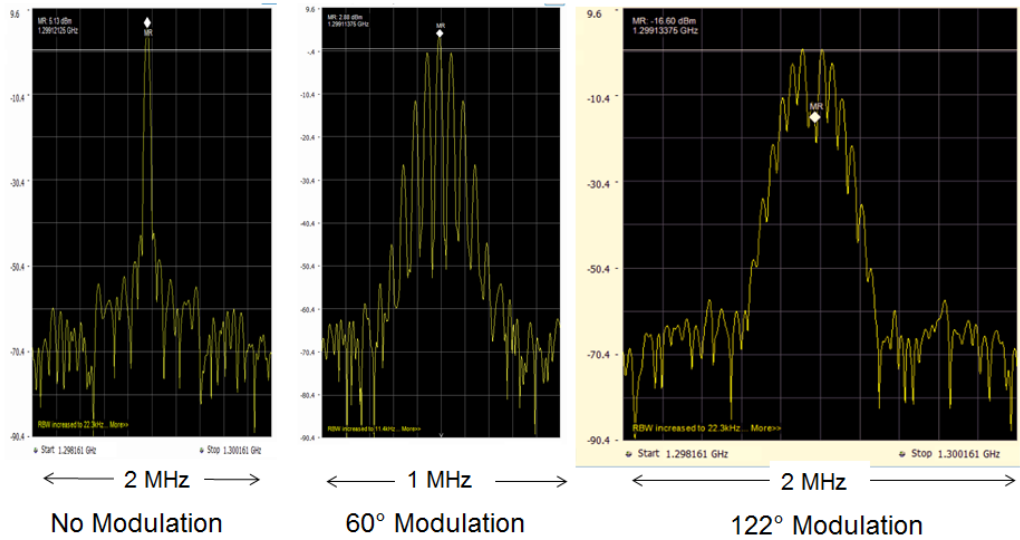

Figure 4. 50 kHz Phase modulation with 269 W locking power and 66.5 kW magnetron output (-25 dB).

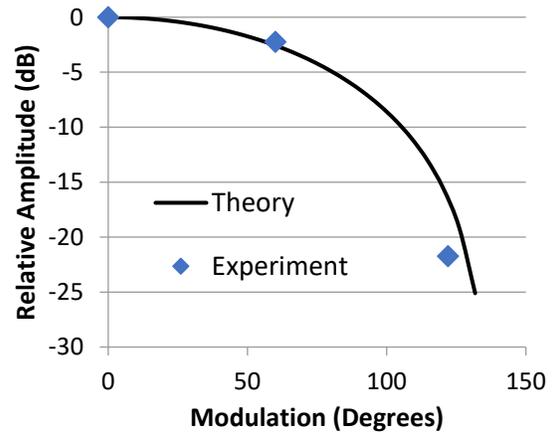

Figure 5 shows the observed power at the center frequency versus the modulation. Also shown is the predicted power, $1-Sii(f,M)^2$, where $Sii(f,M)$ is the amplitude of the side bands and where f is the frequency, $f_c$ is the center frequency, $F_m$ is the modulation frequency, and J0 and Jn are Bessel functions. M is the modulation index, equal to the phase deviation in radians. The measured performance compares well with the predicted values with greater deviation at the highest phase modulation.

Figure 5. Relative power at center frequency as a function of modulation amplitude

$$Sii(f,M) := \left[ J0(M) \cdot \delta(f, f_c) + \sum_{k=1}^{20} \left[ Jn(k,M) \cdot \delta[f, (f_c + k \cdot F_m)] + (-1)^k \cdot Jn(k,M) \cdot \delta[f, (f_c - k \cdot F_m)] \right] \right]$$

## Estimated Cost

The principal components of the system include the magnetron, solid state locking amplifier, 4-port circulator, power supplies, and control electronics. The system shown in Figure 1 included a klystron and its power supplies, which experiments confirmed could be replaced with a solid-state amplifier. The estimated cost for the upgraded system is provided in Table 1. The estimate is for a single system, without the magnetron power supply/modulator. The modulator should cost less than that for an equivalent klystron, since the voltage will be lower. There would be a considerable cost reduction if multiple systems were purchased. For example, for twenty magnetron-based systems, the cost for each would be less than $84,000. This is less than $1/W of delivered RF power.

Table 1. Estimated cost of a 100 kW 1300 MHz magnetron system with amplitude and phase control

| | |
|---|---|
| Magnetron | $72,000 |
| 500 W SS amplifier for locking | $17,000 |
| Circulator w waveguide | $20,000 |
| Controls | $10,000 |
| Packaging | $10,000 |
| TOTAL | $129,000 |

## Application to Accelerators

The interface between the magnetron system and an accelerator will vary with configuration, but some general observations can be made. The primary difference between magnetron and klystron driven systems is the technique for controlling the frequency, phase, and amplitude. For the magnetron driven system, the frequency and phase are controlled by a locking signal as shown in Figure 4. The CCR system includes a high-power amplifier requiring a control signal of approximately 1 W. This can be generated with a standard signal generator, such as the SG380 RF Signal Generator from Stanford Research Systems [8] and a low cost, low noise amplifier. Slow control of the magnetron amplitude can be accomplished by adjusting the voltage or solenoid current. The magnetron voltage and the solenoid current must be adjusted simultaneously to achieve a specified combination of power and frequency.

When driving a superconducting cavity, the amplitude can be rapidly adjusted using phase modulation. Power in the sidebands, outside of the narrow acceptance band of the cavity, will be reflected into the circulator load. As shown in Figure 5, the power in the center frequency is a monotonic function of the phase modulation amplitude. This can be controlled with the SRS signal source through a 0 -1 V DC input signal.

Control of accelerator cavity input power based on beam loading can be accomplished using a feedback loop. Chase *et al.* presented the details of such a loop in a low power demonstration of the technology using a 2450 MHz magnetron at Fermilab [7].

Since the magnetron's efficiency is almost constant from 30% - 100% of the design power, the power can be tuned close to that required for a particular accelerator cavity, with phase modulation producing fast corrections in the amplitude. This differs from a klystron, where operation at significantly less than the optimum power results in decreased efficiency.

## Summary of Magnetron System Development

This research program demonstrated a high efficiency, magnetron system with phase and amplitude control for driving high Q accelerator cavities. The system produced 100 kW at 1.3 GHz with 1.5 ms pulses. A locking bandwidth of 0.9 MHz was obtained with a drive signal 25 dB below the magnetron output power. Consequently, the system can be controlled at full power with a 316 W locking signal, readily available from commercial, solid-state sources.

The demonstrated average power was about 300 W, which was limited only by the available test time. The design was scaled from a 100 kW, 915 MHz device, so there is high confidence that 10 kW of average power can be achieved.

This work was funded by the U.S. Department of Energy under grant number DE-SC0011229 with valuable technical assistance from Richard Zifko at Fermi National Laboratory.

## Multiple Beam Triodes

Triodes are extremely compact sources of electron beams for RF power generation. Operating in Class C, triode-based RF sources can achieve efficiencies approaching 90%. In addition, the cost is extremely low compared to other sources in this frequency range. Estimated cost is approximately $0.50/W, which is less than ¼ the cost of klystrons or solid-state sources.

Figure 6 shows a schematic diagram of a planar triode. A potential difference between the cathode and anode plate drives electron flow when an appropriate voltage is applied to the grid. Because of the small grid to cathode displacement, the current between the cathode and anode is controlled by a relatively small voltage difference between grid and cathode.

Figure 7 shows the basic circuit of a triode amplifier [9,10]. The major components are a generator providing an input RF signal $e_g$, a grid bias voltage source $E_{cc}$, an anode (high voltage) source $E_{bb}$, and a load resistor $R_L$. Because of the load resistor $R_L$, the potential $e_b$ between the plate and the cathode depends on the magnitude of the high voltage source $E_{bb}$ and the magnitude of the anode (beam) current $i_b$. The magnitude of $i_b$ is controlled by the potential between grid and cathode $e_c$. The voltage amplification or gain K is defined as

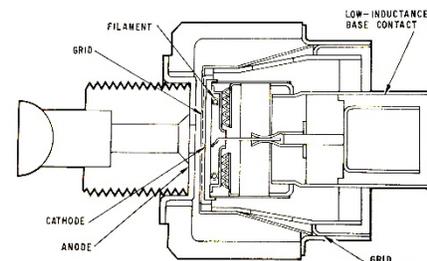

$$K = \frac{\Delta eb}{\Delta ec} = \frac{\Delta ib}{\Delta ec} R_L$$

The advantage of a triode is that a large output power can be generated with a small amount of input power because the grid current is usually negligibly small.

Figure 6. Schematic diagram of a planar triode

The high frequency performance of a triode is limited, in part, by the capacitance between grid and anode. This results in a gain dependent change in the input capacitance known as the Miller effect. This can be avoided by inserting a coarse mesh, referred to as a screen grid, between the control grid and anode. The screen grid shields the control grid from capacitance between the grid and anode. This device is referred to as a tetrode.

For high power gridded tubes with significant beam current, the shielding can partially be achieved by space charge in the electron beam. This research program focused exclusively on the triode.

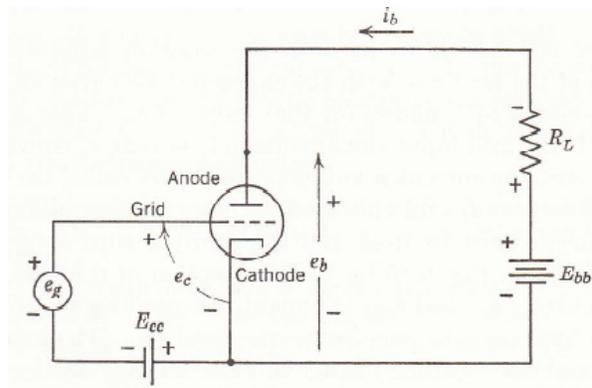

When integrated with input and output cavities, the combination can provide an extremely compact, low cost, high efficiency source of RF power. Figure 8 shows a 425 MHz RF source at CPI used for testing triodes. It produces 25 kW of RF power at an efficiency of 90% and cost approximately $30K. The size can be determined by comparison with the pencil shown in the photo.

Figure 7. Basic circuit of a triode amplifier

This program is developing a multiple beam version of this device to produce 200 kW at 352 MHz [11]. Triode based RF sources consist of a triode that provides the pulsed RF beam, an input cavity that drives the gun grid in Class C, and an output cavity that converts electron beam power into RF power. Consequently, these sources only consist of two coaxial cavities. While efficiencies of 90% can be achieved, they exhibit relatively low gain, with 14 dB being typical. This program is tasked with producing approximately 200 kW of RF power. Consequently, approximately 6 kW of drive power will be required.

The multiple beam triode consists of eight grid-cathode assemblies from a production triode mounted in a circular pattern on the copper support plate. These are driven in parallel by a common input cavity to produce eight electron beams across the grid – anode gap inside a vacuum tube. Figure 9 shows a solid model of the tube. This is the only component in the RF source that's under vacuum.

The multiple beam RF source incorporates an input cavity below the cathodes and an output cavity that extends above and below the grid-anode gap. Because RF power densities are so low, cooling in only required for the vacuum tube anode and grid support plate. The surrounding cavities are primarily fabricated from aluminum. These upper and lower output cavity sections incorporate sliding shorts which determine the inductance of the coaxial sections. The tuners can be adjusted to achieve resonance at the operating frequency using the tube's internal capacitance and circuit stray capacitance. Typical tuning range exceeds 150 MHz, as shown in the HFSS simulation in Figure 10. Consequently, a single design can be used for a multiplicity of applications. Figure 11 shows a solid model of the 200 kW RF source with a six-foot human figure, which demonstrates the compactness of the device.

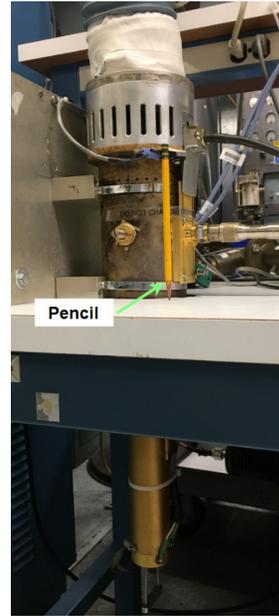

Figure 8. 425 MHz RF source

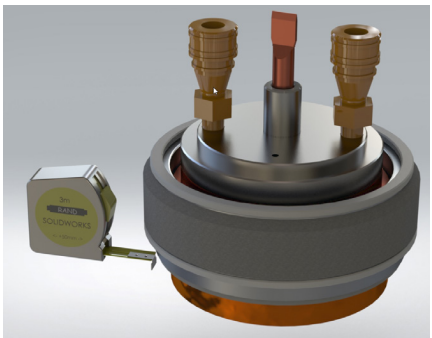

Figure 10. Solid model of multiple beam triode

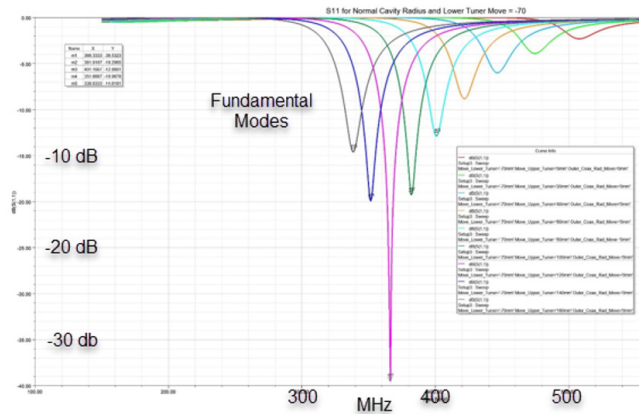

Figure 9. HFSS simulation showing tuning range

One issue with triode base RF sources in the inherent low gain, typically 14 dB. Consequently, a 6 kW RF source will be required to drive the multiple beam tube. This program will use a single beam triode-based RF source provided by JP Accelerator Works as the RF driver. They are a key collaborator in this research program.

## Development Issues

This program is currently dealing with two issues, one electrical and one mechanical. The mechanical issue relates to brazing eight grid-cathode assemblies to a single support structure and ensuring that they are vacuum tight. The program initially encountered materials issues, then design issues, that have prevented successful completion of the assembly. Assembly is in progress with encouraging results. The goal is to obtain a usable assembly for the prototype tube, then revisit the design to optimize the configuration based on experience. The is the most complex and intricate assembly in the device, as eight grid cathode assemblies must be properly integrated into a single, vacuum tight structure. It is anticipated that once this assembly is finished, the vacuum tube can quickly be completed for testing.

The electrical issue relates to coupling of the RF fields to the pulsed electron beam in the grid – anode gap. Significant effort was made to model this structure using HFSS; however, the results were not consistent with experimental results for single beam devices. HFSS predicts lower beam impedance than is consistent with high efficiency. The variation from experience implies that the complex HFSS model is not accurately capturing the configuration, leading to erroneous results.

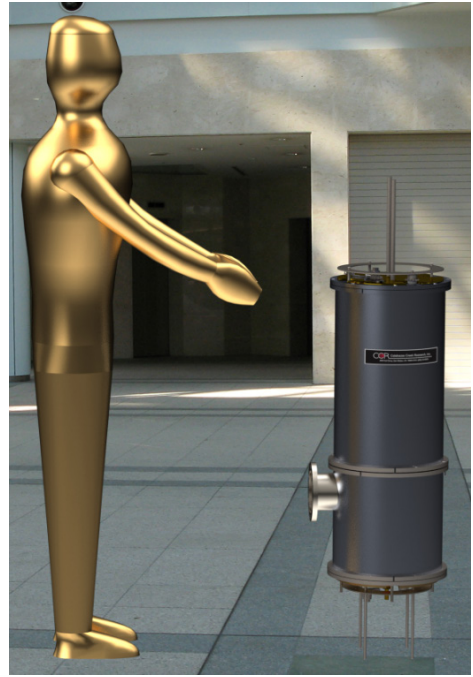

Because of the simplicity of the mechanical construction, the team will build the cavities and experimentally investigate performance. The cavities include tuners and are assembled with fasteners, so modifying the geometry becomes a relatively simple and inexpensive process. This will allow rapid variation of the mechanical configuration without the uncertainties associated with computer modeling. It will also be significantly faster than analyzing with HFSS.

**Figure 11. Image of 200 kW, 350 MHz multiple beam triode RF source model**

There is high confidence that the multiple beam triode source will produce high power, in excess of 100 kW. The question at this point relates to the efficiency that can be achieved.

## Estimated Cost

CCR and CPI began investigating the potential to dramatically increase triode power using multiple beams in 2016. A triode consists of a flat cathode, flat grid, and a flat anode/collector surrounded by a ceramic insulator. There is no magnetic field or focusing optics. The multiple beam device uses the grid cathode assembly from CPI's YU-176 triode. The assembly consists of a nickel cylinder with a barium oxide coating (the cathode), a grid cut with scissors from a tungsten screen, and the appropriate support rings and ceramics. Figure 12 shows a photograph on one assembly. These are in production at CPI and cost less than $1,800 each.

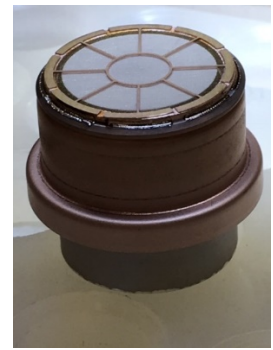

Oxide cathodes are extremely simple and inexpensive; however, they are limited to an average current emission density of 0.25 A/cm$^2$ for CW operation. Even though triodes operate in Class C, the RMS average current emission density of the YU-176 cathode is 1.45 A/cm$^2$ for 25 kW operation. Consequently, the duty is limited to less than 18%. A dispenser cathode for the YU-176 costs approximately $2,800 in quantities of ten or more. One still must add the grid assembly, which can be the same as used with the oxide cathodes.

**Figure 12. Grid-Cathode Assembly**

The parts cost for the multiple beam tube prototype with oxide cathodes was approximately $38K. The addition parts cost for the cavities and other components was approximately $35K. Most labor and assembly cost are associated with the vacuum tube. There's only one braze for the outer cavities, and everything else is welded or attached with fasteners. It's anticipated that the complete 200 kW multiple beam RF source will cost approximately $100K. Adding $35K for the single beam, triode-based driver brings the total cost to approximately $150K, for a cost per Watt of 75 cents. Cost could be significantly less for production quantities.

## High Efficiency Klystron

The U.S. Department of Energy is funding CCR to investigate new approaches to increase the efficiency of klystrons [12]. Specifically, CCR is developing a 100 kW CW, 1.3 GHz klystron with a targeted efficiency exceeding 80%. During the initial research, the program investigated the "Bunch, Align, and Collect" (BAC) method proposed by Gusilov [13] and the "Core Oscillation Method" (COM) proposed by Bajkov 14]. Following extensive analysis of both techniques using AJDISK with optimization and TESLA, it was determined that both approaches could achieve the targeted efficiency, as shown in Figure 13. CCR elected to down select the approach based on mechanical complexity. The BAC approach required fifteen cavities, while the COM approach required eight. As shown in Figure 13, the COM approach requires a circuit approximately 0.5 m longer. After consulting potential users, it was determined that most all facilities could accommodate a longer klystron; consequently, CCR selected the COM design.

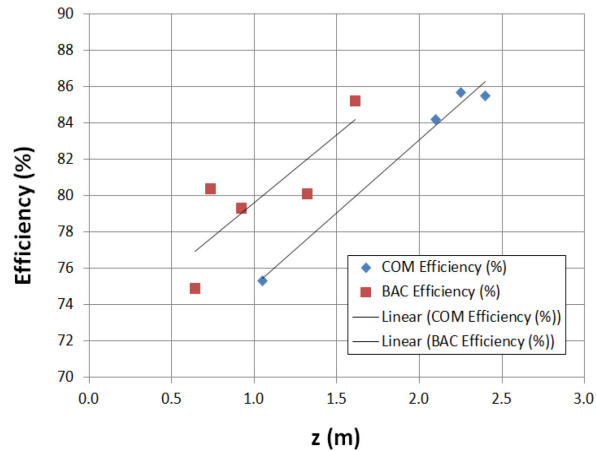

Figure 13. Circuit efficiency as a function of circuit length for both BAC and COM designs. The klystron was simulated using AJDISK. The voltage was 52 kV and the current, 2.3 A for all designs.

The design was simulated using four codes: TESLA, from the Naval Research Laboratory; AJDISK from A. Jensen, KLYC, from CERN, and MAGIC-2D. The results are shown in Table 2. The simulated efficiency agreed within 3%.

Table 2. Results for simulations of the klystron.

| Code | Power | Efficiency |
|---|---|---|
| TESLA | 104.5 kW | 79.5% |
| AJDISK | 106 kW | 81% |
| KLYC | 103.5 kW | 79% |
| MAGIC | 102 kW | 78% |

Table 3 provides the operating parameters, and Figure 14 shows the cross section of the klystron. All components are complete, and final assembly is in progress. The tube is scheduled for bakeout at the end of May 2022. The tube will be baked and tested at CPI, and modification of the test facility is in progress. It is anticipated that testing will begin in early summer 2022.

Table 3. Operating parameters for 1.3 GHz klystron

| Power | 104.4 kW |
|---|---|
| Frequency | 1.3 GHz |
| Voltage | 53.5 kV |
| Current | 2.46 A |
| Efficiency | 79.4% |
| Number of cavities | 7 |

There are two significant challenges related to tube assembly. The first is associated with the cavity frequencies and Qs. Simulations define the tolerances associated with these parameters, which are quite specific. Tuners were incorporated into cavities were the frequency must be particularly precise. Each cavity is cold tested and modified as required to achieve the required performance. As cavities are completed, the computer models are updated with the measured parameters to ensure the target efficiency can be achieved. The input and buncher cavities are complete, and final testing of the output cavity with the output window is in progress.

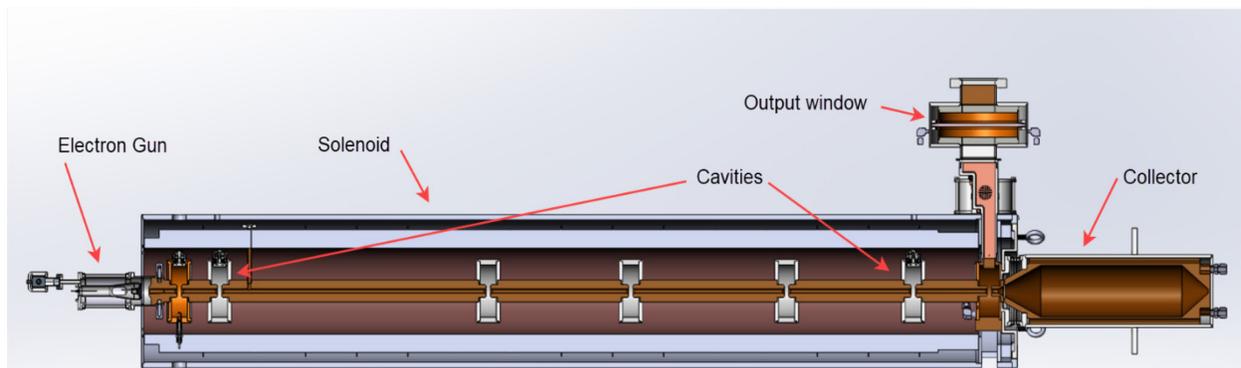

Figure 14. Cross section of the klystron. The length is 2.8 meters.

The second challenge is ensuring the tube remains straight during bakeout, transport, and installation into the test set. The copper beam tunnels are quite long, so the circuit is enclosed within a stainless-steel cage. This cage must accommodate the differential expansion that will occur between the cage and the copper drift tubes. Following bakeout, the tube must be moved horizontally from the assembly building to the test building, a distance of 500 meters with numerous elevation changes.

## Estimated Cost

It's premature to provide an accurate estimate of the cost, since the klystron is fairly complex and assembly is still in progress. The parts costs for the prototype klystron were approximately $115K, and the solenoid cost approximately $64K. The labor hours expended on the prototype klystron are quite high, exceeding 1000 hours, as is typical for any first-time build, and these should drop significantly for future builds. Bakeout, processing, and testing cost will also add to the total. It would not be unreasonable for the total cost to exceed $400K, or $4/Watt for the klystron and magnet.

One task for high power testing will be looking for opportunities to simplify the design or flexibility in parameters, such as cavity tunings, that could lower the cost. In the end, there will probably be a tradeoff between acquisition cost and operating cost. If the klystron efficiency exceeds the goal of 80%, the acquisition cost might be justified by the long-term operating cost, particularly for a CW device.

## Summary of High Efficiency Klystron Development

The goal of this program is to determine the efficiency that can be achieved using advanced circuit design approaches and quantifying the additional fabrication cost, specifically for the COM approach. For applications where a longer tube is acceptable, successful demonstration of high efficiency in the prototype klystron may provide a roadmap for future klystron development. This research program is focused on precisely fabricating the klystron according to the design parameters identified by extensive computer analysis and optimization. The goal is to provide the best opportunity for demonstrating high efficiency operation in a high-power klystron.

## Multiple Beam IOT

A major goal at DOE is development of high efficiency, MW-class, high duty RF sources for several potential applications for RF power. Recent demonstration of 1.2MW multiple beam IOTs (MBIOTs) by CPI-Thales and L3 Electron Devices (L3) demonstrated that these sources can provide this level of power, at least at low duties [15,16]. These devices, though achieving the power and efficiency specifications, were quite complex and expensive. The goal of the CCR program is to investigate techniques for increasing efficiency and reducing the cost. Consequently, the program is focused on three innovations. These include:

- Adding $3^{rd}$ harmonic drive power to improve electron bunching,
- Employing a power splitter approach to simplify the input coupler,
- Replacing pyrolytic graphite grids with moly grids to reduce cost and risk.

The operating parameters are provided in Table 4.

**Table 4. MBIOT operating parameters**

| Operating Frequency | 700 MHz |
|---|---|
| Output Power | 200 kW CW |
| Beam Voltage | 30 kV |
| Total peak current | 6.67A |
| Number of beams | 8 |
| Gain | 23 dB |
| Target efficiency | 85% |

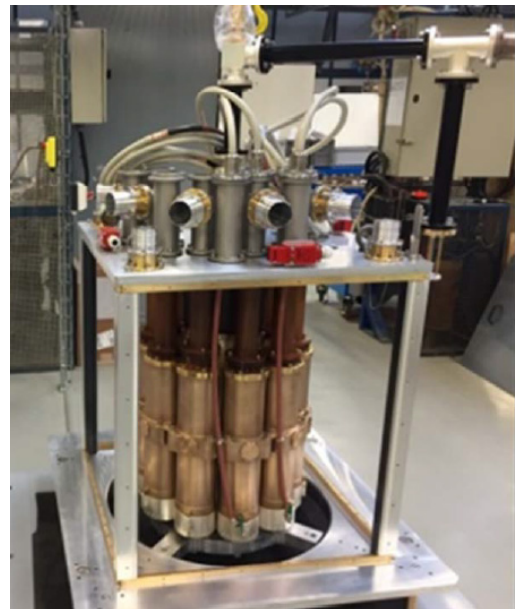

**Figure 15. Input coupler for CPI MBIOT for the European Spallation Source**

The MBIOTs developed by CPI-Thales and L3 employed separate drives and input cavities for each electron gun. For CPI-Thales, this required ten input cavities with drive lines. Consequently, the IOT input section, shown in Figure 15, dominated the device in terms of size and complexity. This complexity was driven by uncertainties concerning the level of amplitude and phase variation that could be present and still provide high power at high efficiency. The separate input couplers allowed variation of drive amplitude and phase for each gun. In the end, experimental measurements determined that the IOT could handle significant variations in amplitude and phase without significant performance degradation. This was confirmed with tests of the L3 MBIOT. This allowed CCR to pursue a simpler input coupler without the complexity of individual gun amplitude and phase controls [17].

Because multiple guns will provide the beam power, CCR investigated replacing traditional pyrolytic graphite (PG) grids with molybdenum grids. Thermal analysis, including consideration of beam loading in Class C operation, confirmed that moly grids could handle the anticipate heating.

It is anticipated that moly grids will be significantly less expensive than PG in addition to being more readily available. While few sources are available for PG grids, thousands of moly grids are produced each year for TWTs and klystrons. This reduces the risk associated with grid availability.

The final innovation is the addition of 3rd harmonic power to the drive signal. Analysis with the advanced simulation code NEMISIS predicts a 2 – 4 % increase in efficiency with the addition of 3rd harmonic drive [18]. Optimum performance is predicted when the 3rd harmonic is approximately 40% of the fundamental and phase shifted by $\pi$ radians. This primarily impacts electrons outside the main bunch created by the fundamental drive.

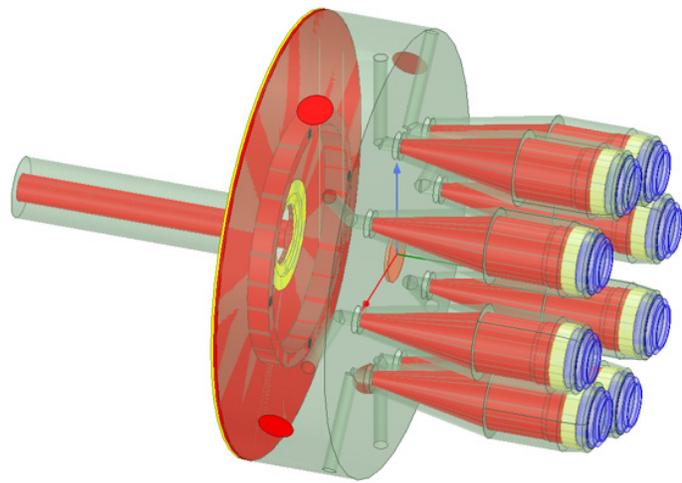

Significant effort was required to develop an input coupler that provided adequate transmission of both the fundamental and the 3rd harmonic. While the 3rd harmonic match is not optimum, it should be adequate to evaluate the impact on MBIOT performance. Figure 16 shows the input coupler geometry interfaced to the eight-gun assemblies. CCR will use a commercial combiner to input the fundamental and 3rd harmonic signals into the input, coaxial transmission line.

**Figure 16. Input coupler geometry**

A key question is whether or not the increased cost and complexity for the 3rd harmonic drive is justified by the efficiency increase achieved. One must consider the cost for an additional driver, combiner, and control system in comparison to the reduced operating cost from higher efficiency.

Performance of the moly grids was simulated using CCR's Beam Optics Analyzer (BOA), which includes 3D, finite element, time domain analysis of the electron beam with complete thermal analysis of the grids. The thermal analysis incorporated the pulsed heating from beam interception and continuous radiant heating from the cathode.

Heating of PG grids is dominated by radiant heating from the cathode due to it's emissivity. The simulated PG temperature is approximately 600º C, while the moly grid temperature is approximately 470ºC. When RF is initiated, the temperature increase for the PG is significantly less that for moly, as is the thermal displacement. The PG grid displacement is approximately 7 µm, while that for moly is approximately 54 µm. The electron gun design compensates for this increased displacement. In all cases, the moly grid remains well below stress limits and temperatures that could result in grid emission.

## Program Status

Design of the electron gun, input coupler, output cavity, output window, and collector are complete. The MBIOT will use a common, cylindrical collector for all eight beams. Figure 17 shows a solid model of the MBIOT. Drawing are in progress and parts procurement will begin as drawings are finalized and released.

## Estimated Cost

Since no parts have been procured, estimating the MBIOT cost is only speculative, at best. The electron gun is based on the K2 gun used in production tubes at CPI. Previous cost was approximately $15K each, suggesting the eight guns for the 700 MHz tube will cost approximately $120K. The high voltage ceramic assembly cost is estimated at approximately $10K, and the coaxial output window cost is estimated to be $20K. The remaining metal parts are fabricated primarily from copper and stainless steel, and none appear to be problematic. Estimated cost for the remaining parts is $50K. The MBIOT will required a solenoid for beam transmission with an estimated cost of $25K. This brings the total parts cost to approximately $225K. The MBIOT should be somewhat easier to assemble, manage, and transport than the klystron, since it's more compact and mechanically robust. A reasonable estimate for the total cost of a tube and magnet would be $350K. For a 200 kW source, this is $1.75/Watt.

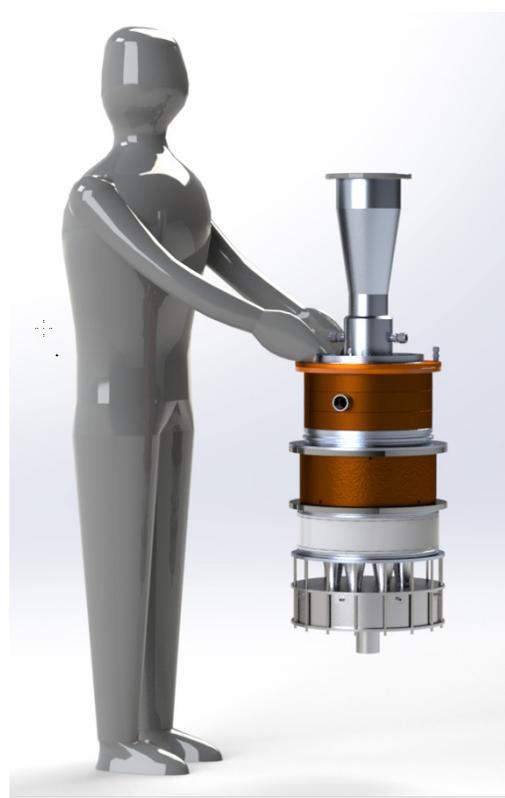

Figure 17. Solid model of 700 MHz, 200 kW MBIOT

## Summary of MBIOT Development

Recent experimental results demonstrated that MBIOTs can provide MW levels of RF power at high efficiency. Fabricating reasonable cost MBIOTs at MW-relevant power levels and high duty remains to be demonstrated and is the principal thrust of this program. Significant cost reduction will be demonstrated if the input coupler design is successfully implemented along with moly grids. The program will also determine the potential impact of $3^{rd}$ harmonic drive on RF performance. Computational design is complete and drawing generation is underway. Parts procurement will begin in the next few weeks, with high power testing of the MBIOT scheduled for early 2023.